\documentclass[conference]{IEEEtran}
\IEEEoverridecommandlockouts
\usepackage{cite}
\usepackage{amsmath,amssymb,amsfonts}
\usepackage{graphicx}
\usepackage{textcomp}
\usepackage{xcolor}
\usepackage{xspace}
\usepackage{xurl}
\def\BibTeX{{\rm B\kern-.05em{\sc i\kern-.025em b}\kern-.08em
    T\kern-.1667em\lower.7ex\hbox{E}\kern-.125emX}}
\usepackage{mathtools}
\usepackage{multirow}
\usepackage{adjustbox}
\usepackage{tikz}
\usetikzlibrary{positioning}
\usepackage{hyperref}
\usepackage{caption}
\usepackage[french,boxed,vlined,linesnumbered,inoutnumbered,rightnl,algo2e]{algorithm2e}
\usepackage{algorithm}
\usepackage{graphicx}
\usepackage{mwe}
\usepackage{stfloats}
\usepackage{hyperref}
\usepackage{threeparttable}
\begin{document}

\title{
Towards Auto-Building of Embedded FPGA-based Soft Sensors for Wastewater Flow Estimation
\thanks{The authors gratefully acknowledge the financial support provided by the Federal Ministry for Economic Affairs and Climate Action of Germany for the RIWWER project (01MD22007C).}}
\author{
 \IEEEauthorblockN{Tianheng Ling, Chao Qian, Gregor Schiele}
 \IEEEauthorblockA{Department of Intelligent Embedded Systems, University of Duisburg-Essen, Duisburg, Germany}
 \IEEEauthorblockA{\{tianheng.ling, chao.qian, gregor.schiele@uni-due.de\}}}
\maketitle
\begin{abstract}

Executing flow estimation using Deep Learning (DL)-based soft sensors on resource-limited IoT devices has demonstrated promise in terms of reliability and energy efficiency. However, its application in the field of wastewater flow estimation remains underexplored due to: (1) a lack of available datasets, (2) inconvenient toolchains for on-device AI model development and deployment, and (3) hardware platforms designed for general DL purposes rather than being optimized for energy-efficient soft sensor applications. This study addresses these gaps by proposing an automated, end-to-end solution for wastewater flow estimation using a prototype IoT device.

\end{abstract}


\section{Introduction and Related Work}

Combined sewer systems (CSS) are widely used worldwide but face increasing challenges from extreme weather exacerbated by pollution and climate change. These conditions often lead to urban overflow events, significantly impacting public health and environmental sustainability~\cite{ahm2016estimation}. Real-time control systems offer a promising alternative by preventing CSS overflow through dynamic flow monitoring. However, these systems are hindered by high operational costs and challenges in achieving continuous, accurate monitoring and real-time data processing~\cite{sun2020integrated}. Current flow monitoring techniques mainly rely on hydrological formulas, which, while cost-effective, lack the precision necessary for effective management. Advanced sensor technologies, such as Coriolis and Magnetic Induction devices, offer higher accuracy but are cost-prohibitive for widespread adoption in economically constrained regions~\cite{noori2020non}.

Previous research has demonstrated the promising potential of soft sensors for flow estimation based on fluid level~\cite{ahm2016estimation,noori2020non}. Specifically, Noori et al.~\cite{noori2020non} simulated drilling conditions using a flow circulation system with a Venturi structure, utilizing three level sensors for input data, while a Coriolis mass flow meter provided target data. A simple Multi-Layer Perceptron (MLP) model was then used to efficiently estimate non-Newtonian fluid flow.

Building on these datasets, our prior work~\cite{ling2023device} implemented a quantized MLP model on a low-power MCU (Cortex-M0 RP2040) and an embedded FPGA (Xilinx Spartan-7 XC7S15) using TensorFlow Lite and our customized ElasticAI.Creator toolchains~\cite{qian2023elasticai}, respectively. The FPGA-deployed models demonstrated up to 28.44$\times$ faster inference speed compared to MCU-deployed models, easily meeting real-time requirements. However, due to differences in applied quantization technologies, FPGA-deployed models exhibited lower precision. By applying integer-only quantization, we increased the model precision by up to 9.7\%~\cite{ling2024flow}. Additionally, implementing pipelined matrix multiplication on the FPGA reduced inference time by up to 9.39\%, with an energy cost per inference only up to 5.71\% higher.

\begin{figure}[]
    \centering
    \includegraphics[width=.9\columnwidth]{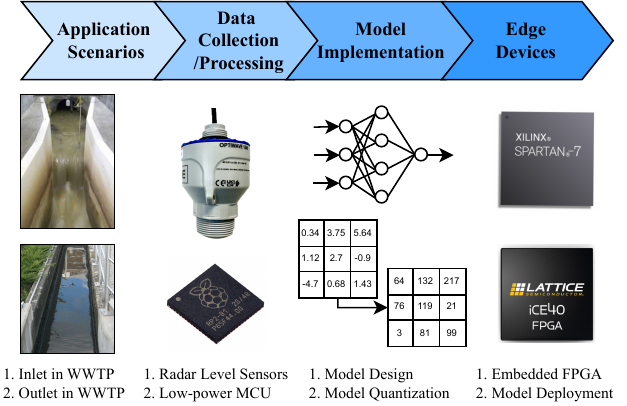}
    \caption{Workflow of Flow Estimation with Soft Sensors from Application to Edge Devices}
    \label{fig:system_architecture}
    \vspace{-15pt}
\end{figure}

Despite these advancements, there remains a need for datasets specifically for wastewater flow estimation. Additionally, the deployment of an integer-only quantized MLP model is not automated, making it challenging for researchers without FPGA expertise. Moreover, the previous hardware platform was overly powerful for this soft sensor application, leading to inefficient resource utilization and significant power consumption, resulting in energy inefficiency. This research explores end-to-end optimization strategies from sensor integration to model deployment, as shown in Figure \ref{fig:system_architecture}\footnote{Some images are sourced from the internet, with all rights reserved by the original authors.}. The goal is to develop a scalable, efficient system capable of real-time, accurate wastewater flow measurement, thereby enhancing in-field monitoring capabilities in managing urban overflows.

\vspace{-5pt}
\section{Expected Contribution}

The expected contributions of this research include collecting wastewater flow data with customized test bench, automating model deployment on embedded FPGAs, and optimizing hardware design to enhance energy efficiency for on-device flow estimation. These contributions are detailed as follows:

\subsection{Application Scenarios-based Data Collection}

Given the inherent differences between wastewater and drilling fluid, we aim to collect relevant data specifically for wastewater flow estimation. The primary experimental setups include open channel flumes, such as those found at the inlet and outlet of wastewater treatment plants (WWTP), as illustrated in Figure \ref{fig:system_architecture}. 

\begin{figure}[!htb]
    \vspace{-8pt}
    \centering
    \includegraphics[width=.74\columnwidth]{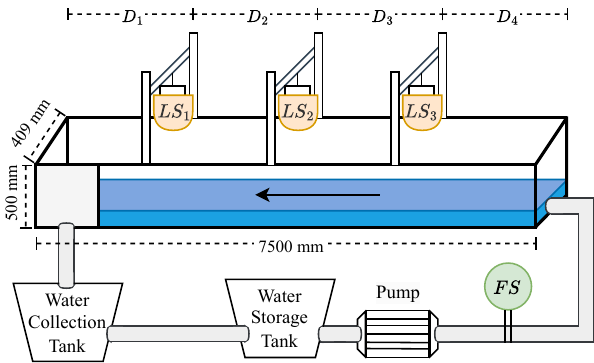}
    \caption{Illustration of GUNT HM162 Experimental Flume}
    \label{fig:flow_loop}
    \vspace{-5pt}
\end{figure}

Experiments will be conducted using the GUNT HM 162 experimental flume, as simplified in Figure \ref{fig:flow_loop}. In this setup, a pump transfers water from a storage tank into the flume and back. The flume is equipped with a flow sensor (\(FS\)) to collect flow data. Three adjustable brackets above the flume are used to hang KROHNE Optiwave radar level sensors (\(LS_{1}\) to \(LS_{3}\)). The positions of these sensors are indicated by \(D_{1}\) to \(D_{4}\) and are adjustable. The flume's slope can vary from -0.5\% to +2.5\%, allowing simulation of various flow conditions and identification of the optimal settings for precise flow estimation. Additional structures, such as Venturi, Paschal, and trapezoidal flumes, can be installed to examine different hydraulic scenarios. We will also adjust the number of radar sensors used to achieve accurate flow estimation with minimal sensors. Additionally, we will simulate the characteristics of wastewater by adding bubbles, solids, and other contaminants to match real-world conditions.


\subsection{Automated Model Implementation}

Our open-source ElasticAI.Creator\footnote{\url{https://github.com/es-ude/elastic-ai.creator}} toolchain is expected to be advanced to automate the generation of accelerators for soft sensors. To achieve this, we first integrate integer-only quantized MLP models into ElasticAI.Creator. Using this toolchain, we develop and train the MLP models with the collected wastewater datasets and subsequently generate the corresponding accelerators. Our objective is to enable users to simply prepare a dataset and, with a single command, produce an executable accelerator suitable for deployment on our supported IoT devices. This streamlined process aims to facilitate the deployment of DL models on embedded FPGAs, making it accessible to those without FPGA expertise.

\subsection{Optimized Hardware System Design}

\begin{figure}[!htb]
  \centering
  \begin{minipage}{0.32\columnwidth}
    \centering
    \includegraphics[width=1\columnwidth]{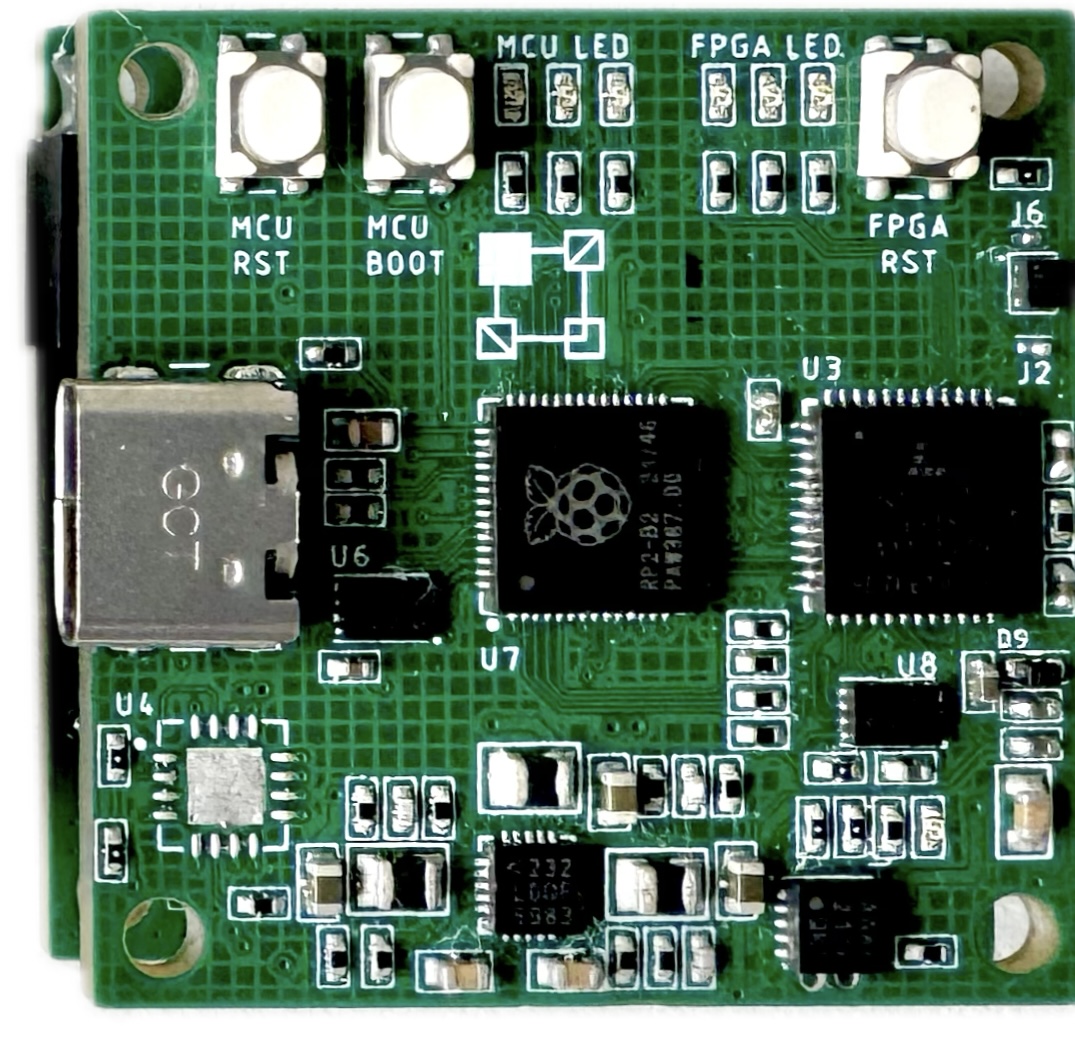}
  \end{minipage}
  \hspace{5pt}
  \begin{minipage}{0.5\columnwidth}
    \centering
    \includegraphics[width=1\columnwidth]{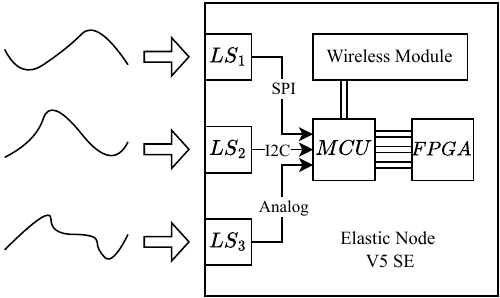}
  \end{minipage}
\caption{Elastic Node V5 SE with Soft Sensor}
\label{fig:hardware}
\vspace{-10pt}
\end{figure}

In this study, we also propose an optimized hardware design targeting energy-efficient soft sensor applications. The modification involves replacing the XC7S15 FPGA with a smaller FPGA, the ICE40UP5K. This decision is based on our previous work~\cite{ling2024flow}, which indicated low resource utilization of our MLP accelerator on the XC7S15 FPGA (6.47\% LUTs, 7.5\% BRAMs, and 10\% DSPs). This low utilization suggests that the design can be shifted to a smaller FPGA without performance loss. Additionally, the XC7S15 FPGA has significant static power consumption (approximately 30 mW), whereas the ICE40UP5K’s static power consumption is at the micro-watt level. Thus, this replacement could markedly improve the system’s energy efficiency. To address the challenges posed by resource constraints, we plan to implement advanced hardware optimizations in the accelerator design. These optimizations will ensure that the design remains efficient and effective despite the reduced resources available on the ICE40UP5K.

\section{Conclusion}

This study explores the application of DL-based soft sensors for real-time, precise wastewater flow estimation with embedded FPGA and radar sensors. Our research focuses on optimizing the entire process, from sensor integration to model deployment, automating accelerator generation, and enhancing hardware design for better energy efficiency and performance. By addressing these aspects, we aim to improve the adaptability and effectiveness of flow monitoring systems in urban environments.

\bibliographystyle{IEEEtran}
\bibliography{reference}
\end{document}